\documentclass[10pt]{article}
\usepackage{authblk}
\usepackage{amsmath}
\usepackage{tabularx}
\usepackage{graphics}

\title{Thermo-optic coefficient measurement of liquids using silicon photonic microring resonators}

\author[1,2]{Prashanth R Prasad}
\author[1]{Shankar K Selvaraja}
\author[1,2]{Manoj Varma\thanks{mvarma@iisc.ac.in}}

\affil[1]{\small{Centre for Nano Science and Engineering, Indian Institute of Science, Bangalore. }}
\affil[2]{\small{Department of Electrical Communication Engineering, Indian Institute of Science, Bangalore.}}

\date{}
\begin{document}
\maketitle

\begin{abstract}
On-chip measurement of thermo-optic coefficient (TOC) of samples along with on-chip temperature measurements can be used to compensate
for thermal fluctuation induced noise in refractometry using integrated photonic sensors. In this article we demonstrate a device
design and describe the method to extract TOCs of liquid samples using resonant wavelength shifts of a silicon microring resonator.
The TOCs of three standard fluids; De-ionized water, Ethanol and Isopropanol, were determined using our sensor and show a good
agreement with values reported in literature. A mechanism for tracking of on-chip temperature variations is also included for ensuring
accuracy of TOC measurements. Potential applications of the demonstrated on-chip TOC sensor include improvements in accuracy of
refractive index measurements and multiparametric analysis of biochemical analytes.
\end{abstract}

\section{Introduction}

Silicon photonic sensors have been studied extensively for measurement of refractive index of liquids. Their unique advantages such
as small size and the potential for large-scale manufacture using CMOS compatible technologies have made them
promising candidates for applications in several areas including healthcare, biomedical research and
environmental monitoring \cite{Estevez2012}. Among various implementations of Silicon photonic sensors, microring resonators have been
investigated widely because of their
compactness \cite{Xu2008}, high quality factors \cite{Niehusmann2004} and scalability for multi-analyte sensing using arrays of sensor elements \cite{Vos2007,Vos2009}.
An important figure-of-merit of the Silicon photonic sensor is its detection limit, or the smallest change of refractive index that
can be reliably measured by the system. 
In the past, highly sensitive Silicon microring sensors capable of measuring index shifts with a detection limit of $10^{-7}$
Refractive Index Units (RIU) have been reported \cite{Iqbal2010}. 
An important factor that influences the detection limit of the silicon microring sensor is the variation in operating temperature of the device.  
Owing to the high thermo-optic coefficient (TOC) of silicon ($1.86\times10^{-4}/\,^{\circ}\mathrm{C}$) \cite{Frey2006}, 
small amplitude temperature fluctuations in the device can cause significant but undesirable shifts of resonances in a
microring sensor \cite{Bogaerts2014}. 
Usually, active controller platforms such as a peltier thermo-electric coolers (TECs) are employed to compensate for random
fluctuations in the ambient temperature. However, it is difficult to track and
correct localized thermal variations on the chip surface because of the large response times of closed loop controllers. Such
variations are possible due to differences in temperatures of the sensor chip surface and the analyte fluid introduced from an
external reservoir. In the past, reference microring resonators have been employed for cancellation of temperature fluctuations at the
surface of the sensor \cite{Xu2010a,Stern2017}. 
In this approach, wavelength shifts due to on-chip temperature variations are tracked using a dedicated reference microring unexposed
to the analyte being probed. Subtraction of the wavelength shift of reference microring from that of
the sensor ring gives the required signal shift caused by a change in refractive index of analyte. However, this method suffers from a
few drawbacks. Firstly, imperfections in fabrication processes can result in differences between dimensions of waveguides forming
reference and sensor microrings, causing an imbalance in the spectral responses to temperature shifts.  Furthermore, the spatial
separation between the two microrings can result in a difference in local temperatures. These shortcomings can potentially
introduce errors in refractive index calculations. 
On the other hand, the knowledge of thermo-optic coefficient of analyte can directly be used for correction of errors in 
refractive index measurements caused by temperature fluctuations. This process also requires a real time measurement of 
temperature changes in vicinity of the sensor microring. In this article, we demonstrate a method to measure the thermo-optic
coefficient of analyte by using resonance shifts of microrings caused by controlled increase in sensor temperature.
Additionally, an on-chip tracking system is also implemented to monitor temperature changes at the chip-surface in real time.
Besides corrections of errors in refractive index measurements, the TOC of a liquid can also be used as an independent parameter
for chemical analysis. This is because, the TOC of a material is not directly correlated with the refractive index but depends on
various other factors including volumetric expansion coefficient and chemical compositions \cite{Coppola2011,Ghosh1998book,Cao2015}.

In the past, various device configurations such as fibre bragg gratings \cite{Lee2010,Kamikawachi2008}, Fabry-Perot interferometers
\cite{Cao2015}, Hollow core fibres \cite{Lee2015} and two-mode fibre interference \cite{Kim2012} have been used for measurement of
TOCs of liquids. These implementations require
large volumes of analytes for measurements and are therefore not readily adaptable for parallel analysis of multiple liquid
analytes. We have measured the TOCs of standard fluids such as De-ionized water, Ethanol and Isopropanol and compared our results with
values reported in literature. Using our method, the TOC of De-ionized water was determined 
as $-1.12\times10^{-4}/\,^{\circ}\mathrm{C}$ with an estimated error limits of  $\pm8.26\times10^{-6}/\,^{\circ}\mathrm{C}$.

\section{Design and Simulations} \label{sec-design-j3}
We determine the TOC of a liquid by measuring temperature dependent shifts in resonant wavelengths of a silicon microring. Because of the thermo-optic
effect, changes of temperature results in variation
of refractive indices of the waveguide core, liquid-cladding and the buried oxide, causing a net change in the effective index of
the microring waveguide. Using the theory developed for athermal operation of silicon photonic devices \cite{Raghunathan2010}, one can
express the net effective index change (per unit variation in temperature) in terms of shifts in refractive indices of core and
cladding regions as,
\begin{equation}\label{eq-nefftoc}
\frac{\delta n_{eff}}{\delta T} = \Gamma_{core}\frac{\partial n_{core}}{\partial T}~+~\Gamma_{box}\frac{\partial n_{box}}{\partial T}~+~
\Gamma_{fl}\frac{\partial n_{fl}}{\partial T}
\end{equation}
Here, $\Gamma$ is the confinement factor in core, cladding or the buried oxide while $\frac{dn_i}{dT}$ are the thermo-optic
coefficients. The confinement factor $\Gamma$ is defined as
\cite{Robinson2008},
\begin{equation}\label{eq-confinement}
\Gamma_s = \frac{n_g}{n_s}\left( \frac{\iint_s\!\epsilon \! \left|E\right|^2 \text{dxdy}}{\iint_{tot}\epsilon\!\left|E\right|^2 \text{dxdy}}\right) = \frac{n_g}{n_s}\gamma_s 
\end{equation}
In this expression, $n_s$ is the refractive index of the waveguide constituent (core/cladding/oxide) under consideration, $\gamma_s$ is the fraction of energy density within the same region and is termed as `fill factor'.  Presence of group index $n_g$ in
Eq.~\eqref{eq-confinement}
indicates that confinement factors depends on the group velocity of propagating mode. To determine the TOC of analyte
cladding $\left(\frac{\partial n_{fl}}{\partial T}\right)$, we rewrite the equation as,
\begin{equation}\label{eq-tocfluid}
\frac{\partial n_{fl}}{\partial T} = \frac{\frac{\delta n_{eff}}{\delta T}~-~\Gamma_{core}\frac{\partial n_{core}}{\partial T}
	~-~\Gamma_{box}\frac{\partial n_{box}}{\partial T}}{\Gamma_{fl}}
\end{equation}
The rate of change of effective index $\left(\frac{\delta n_{eff}}{\delta T}\right)$ is obtained by measurement of resonant
wavelength shifts  $\left(\frac{d\lambda}{dT}\right)$, and substitution of results in the following expression: 
\begin{equation}\label{eq-dneffdT}
\frac{\partial n_{eff}}{\partial T} = \frac{n_g}{\lambda_{res}} \frac{\Delta \lambda}{\Delta T} 
\end{equation}
where $\lambda_{res}$ is the resonant wavelength.  

\begin{figure}[!htb]
	\begin{center}
		\resizebox{120mm}{!}{\includegraphics *{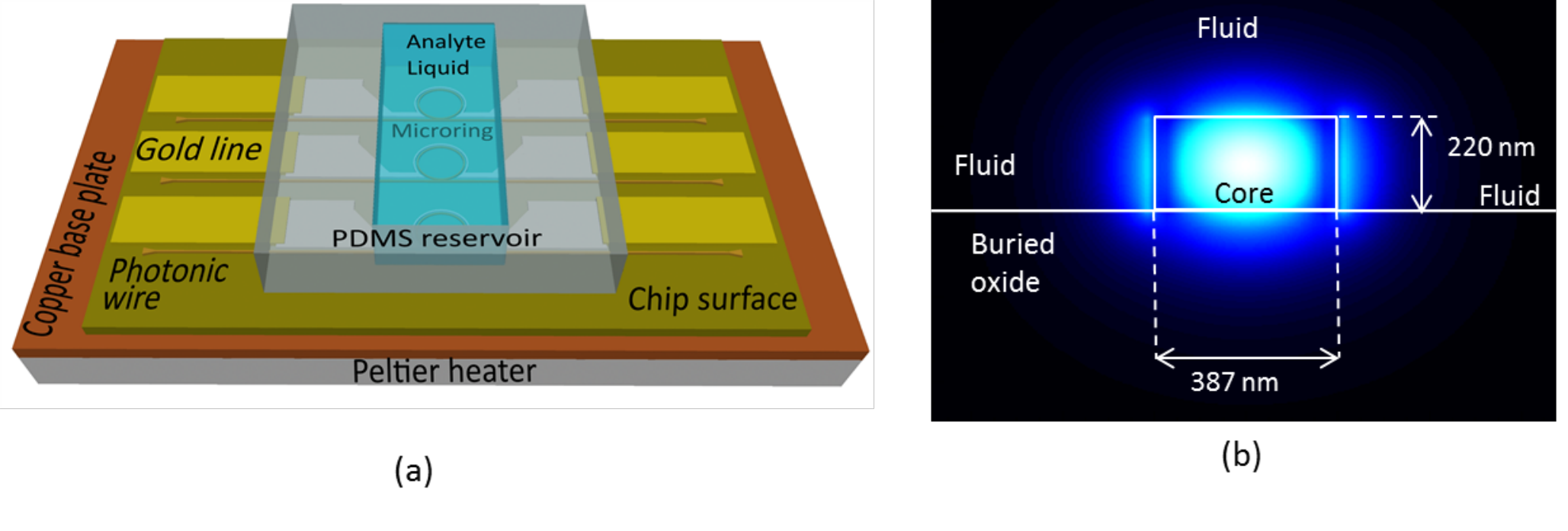}}
		\caption{\small{Design and simulation of silicon microring TOC sensor (a) A schematic showing the sensor system. (b) Simulation result showing the distribution of
		electric field (TE-like mode) along the cross section of a photonic wire waveguide.  }}
		\label{schematic-j3}
	\end{center}
\end{figure}

Schematic of the system used for wavelength shift measurements is shown in
Fig.~\ref{schematic-j3}a. The photonic chip consists of several
identical silicon microring resonators for redundancy. Metal rings are patterned around each resonator, at a distance of about 2$\mu \text{m}$,  for measurement of temperature variations. Low resistance Gold contact lines are defined for electrical probing of 
metal rings. A PDMS (Polydimethylsiloxane) reservoir structure provides containment for liquid analyte. The chip is mounted on a
copper base plate, which itself
is placed on a peltier module. A resistance temperature detector (RTD) sensor is attached to the copper base plate and interfaced to the
temperature controller. This device provides variable electrical power to the peltier heater module for closed-loop control of chip temperature. 
Waveguides and microrings were implemented using photonic wires ($400~\text{nm}\times~300~\text{nm}$) operating with fundamental TE-like mode at
1550 nm wavelength band. Since the core material silicon has a linear expansion coefficient of $2.59\times10^{-6}/\,^{\circ}\mathrm{C}$,
change in waveguide dimensions over the relatively small operating temperature range (about 12 $\,^{\circ}\mathrm{C}$) 
is not considered in our analysis.

Thermo-optic coefficient of silicon $\left(\frac{\partial n_{core}}{\partial T}\right)$ is calculated via resonance shift measurements
and cross-checked with previously reported values, while the TOC of buried oxide $\left(\frac{\partial n_{box}}{\partial T}\right)$ is taken as
$8.57\times10^{-6}/\,^{\circ}\mathrm{C}$ from literature reports \cite{Leviton2008}. 
Confinement factors $\Gamma_i$ are computed by determining fill factors $\gamma_i$ using modal simulations since we have accurate knowledge
of the dimensions and refractive indices of core, fluid and buried oxide regions. Specifically, we used finite element method modal analysis software for computing fill factors.  
 Width and height of the simulated waveguide structure were set to 387 nm and 220 nm
respectively after analyzing the scanning electron microscope (SEM) images of the fabricated device. 
Thickness of the buried oxide layer was set to
2 $\mu m$, based on wafer specifications. Fig.~\ref{schematic-j3}b shows the field
distribution in the waveguide. 
 Refractive index of tested analyte liquids (at 1550 nm), forming the upper and side clads were
taken from literature \cite{Saunders2016}. Furthermore, we have measured the refractive index of the analytes at 590 nm wavelength using a 
commercial Abbe refractometer for comparison.
Tab.~\ref{tab-fillfactors} shows the simulated fill factors for air and analyte claddings. 
\begin{table}[!htb]
	\begin{center}
		\begin{tabularx}{\textwidth}{|l|X|X|l|l|l|l|l|}
			\hline
			Fluid&Ref. Index (1550 nm) &Ref. Index (590 nm)&$\gamma_{core}$&$\gamma_{fl}$ &$\gamma_{box}$&$n_{eff}$&$n_g$\\
			\hline\hline
			Air&1&1&0.727&0.113 &0.146&2.129&4.64\\ \hline
			DI Water&1.3164 \cite{Saunders2016}&1.333&0.71&0.142&0.132&2.208&4.37\\ \hline
			Ethanol&1.3503 \cite{Saunders2016}&1.359&0.708&0.145 &0.13&2.217&4.34\\ \hline
			Isopropanol&1.3661 \cite{Saunders2016}&1.373&0.707&0.147 &0.129&2.221&4.33\\ 
			\hline		
		\end{tabularx}
		\caption{\small{Simulation of confinement factors for different claddings of the photonic wire. Waveguide cross section $387~ \text{nm}\times220~\text{nm}$. Working wavelength is 1510 nm.}}
		\label{tab-fillfactors}
	\end{center}
\end{table}

\section{Experiments and Results}
\begin{figure}[!htb]
	\begin{center}
		\resizebox{100mm}{!}{\includegraphics *{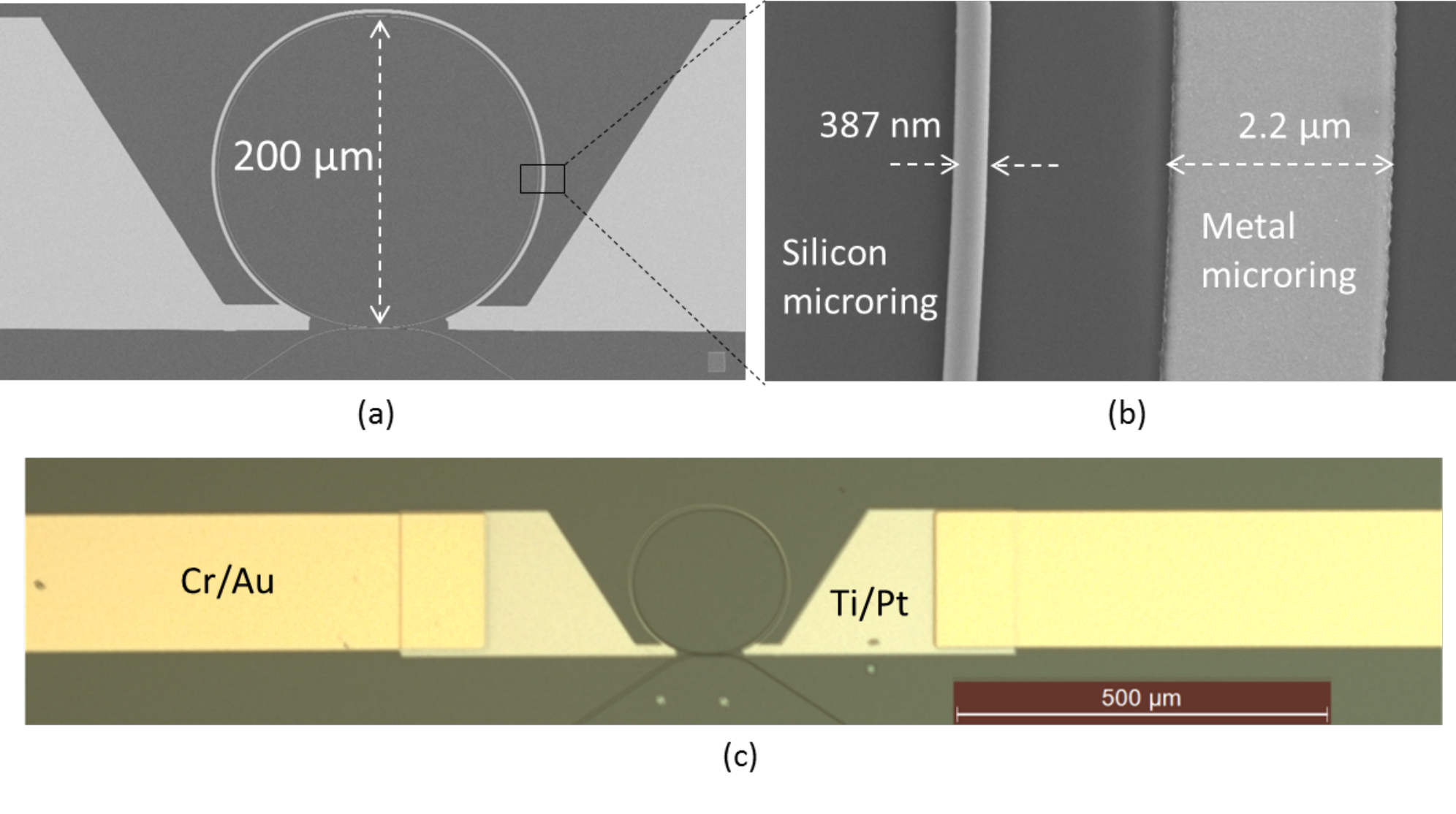}}
		\caption{\small{Images of fabricated microrings and metal rings for thermo-optic coefficient measurement. (a) Scanning Electron
		Microscope (SEM) image of Silicon microring surrounded by platinum metal ring (b) Magnified view of the wire waveguide and
		metal ring (c) Microscope image showing gold contact lines for electrical probing of metal rings}}
		\label{fabimages-j3}
	\end{center}
\end{figure}

An SOI wafer with 220 nm thick device layer
and 2$\mu \text{m}$ buried oxide was used for fabrication of the sensor chip. In the first step, Electron beam (e-beam) lithography was performed
using a negative tone resist (ma-N2401) for definition of photonic wires, microrings and bases for grating couplers. 
Radius of microrings was designed to be 100 $\mu m$, and the width of metal ring was set to 2 $\mu m$. 
A dry etch step using
fluorine plasma was used to transfer resist patterns to the SOI device layer. Following this, a second e-beam lithography process was carried out
using a positive tone (PMMA) resist coating for patterning gratings over previously formed bases. These patterns were transfered onto 
SOI layer via dry etching.
 A third step of  e-beam exposure process (with PMMA resist) was employed for definition
of metal ring patterns. This was followed by sputtering of metal bi-layer (Titanium-10 nm/ Platinum 50 nm) and lift-off processes
to obtain metal structures. 
Subsequently, optical lithography and metal lift off (Chromium-10 nm/ Gold-90 nm) steps were used for definition
of low resistance contact strips for electrical probing. A positive tone resist (S1813) was used for this purpose.
Finally, a containment structure prepared using PDMS was bonded over the sensor chip to serve as a reservoir
for analyte liquid over microrings.  Fig.~\ref{fabimages-j3} shows images of patterns at various stages of fabrication. 

An optical source-detection module 
consisting of a tunable laser (1505-1620 nm range), and a InGaAs power meter was used for spectrum shift measurements. Grating couplers
fabricated on chip were used to couple light in and out of the photonic sensor chip through single mode (SMF-28) optical fibres. A high
precision source-measurement unit was used to monitor the electrical resistance of metal rings and thereby track the on-chip temperature
variations. 

\begin{figure}[!htb]
	\begin{center}
		\resizebox{110mm}{!}{\includegraphics *{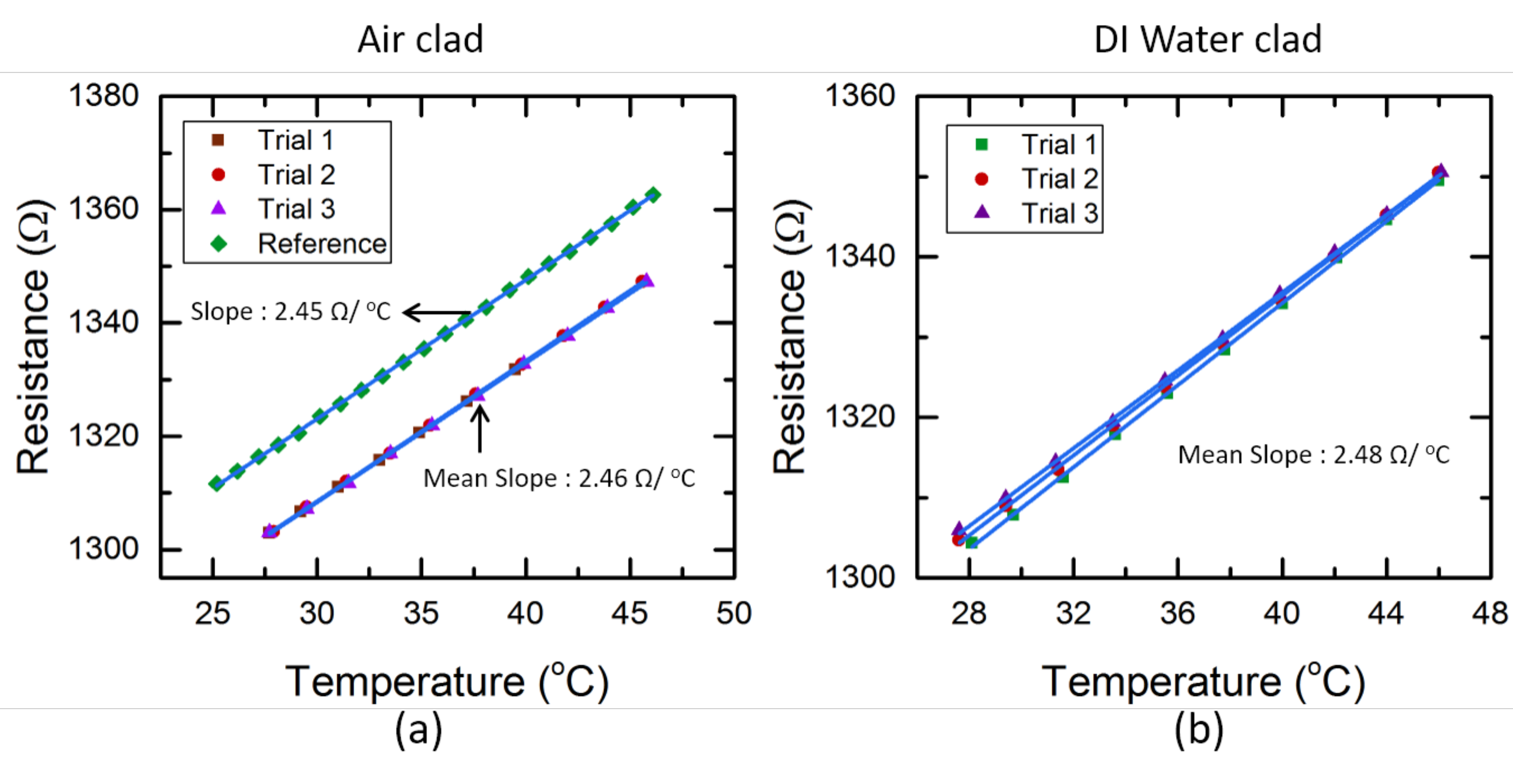}}
		\caption{\small{Measurements of metal ring resistance versus temperature (a) With air covered microrings. The reference data
				was obtained using a calibrated thermal chuck. Three trials were performed using our characterization assembly 
				(b) With DI-water covered microrings. The $R^2$ value exceeded 0.99 for all linear fits.}}
		\label{TCR-Char}
	\end{center}
\end{figure}

The on-chip temperature tracking system comprising the resistive metal ring structure was first calibrated using a device probe station
equipped with a high precision thermal chuck and a semiconductor device analyzer. In this experiment, the electrical resistance of metal ring was monitored for increasing
temperatures of the chuck. 
The resistance of metal ring was measured to be around 1300 $\Omega$ while that of the contact pads 
used for electrical probing was in the range of a few ohms. 
From the
resistance-temperature plot, we extracted the slope of resistance variation to be 2.45 $\Omega/\,^{\circ}\mathrm{C}$.
 For this experiment, the
PDMS reservoir was not filled with any liquid.
Next, 
we performed the same experiment on the experimental assembly described in Sec.~\ref{sec-design-j3} where the peltier heater
was actuated using the controller system to heat the photonic chip to the desired temperature. A source measurement unit (SMU)
was used to monitor the electrical resistance of the metal ring. In this case, the mean slope (3 trials) of resistance variation with temperature was measured to be 2.48 $\Omega/\,^{\circ}\mathrm{C}$, as shown in Fig.~\ref{TCR-Char}a.
This experiment was repeated after filling the PDMS reservoir with DI-water. We ensured that the voltage applied 
to the metal rings for electrical resistance of metal rings was kept below the electrolytic potential of water (1.23V) to 
avoid electrolysis. The mean resistance slope in this case was 2.46 $\Omega/\,^{\circ}\mathrm{C}$, as seen in Fig.~\ref{TCR-Char}b.
Evidently, the thermal
mass of small volume of water (few microliters) does not cause significant differences in the temperature variation relative to
that of air clad. Thus, by monitoring variations in electrical resistance of metal rings, one can back-calculate the temperature
change in vicinity of microring resonator. 

\begin{figure}[!htb]
	\begin{center}
		\resizebox{110mm}{!}{\includegraphics *{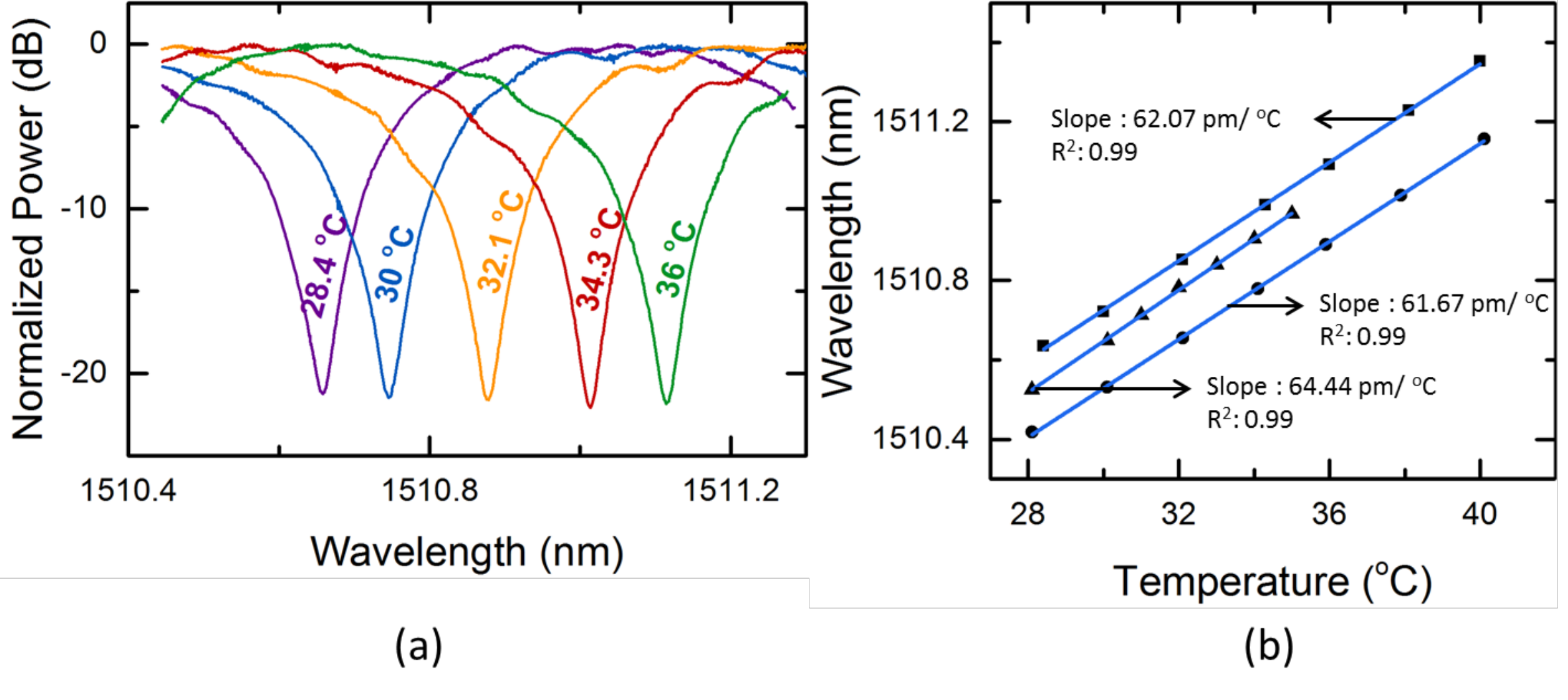}}
		\caption{\small{Measurements for determination of TOC: Silicon (a) Positions of a transmission resonance for different temperatures.
				Curve fitting was used to determine resonance minimum precisely. (b) Shift in resonance position as a function of 	temperature (3 trials). $R^2$ value for all curve fits exceeded 0.99.}}
		\label{TOCsil-j3}
	\end{center}
\end{figure}

We have used the theory described in the previous section to determine the TOC of Silicon. A good agreement of our measurements with 
literature reports would validate the described TOC measurement method. The PDMS reservoir was left unfilled so that air forms the upper and side claddings of the waveguide.
 Temperature of the photonic chip was varied using the closed loop controller while resonant wavelength shifts were monitored.
Measured full width half maximum (FWHM) width of resonances was about 0.1 nm, with an FSR of 0.79 nm.
A linear fit was used to express the wavelength shift as a function of the chip 
temperature (Fig.~\ref{TOCsil-j3}a). Mean slope ($\frac{\Delta \lambda}{\Delta T} $) of three trials was calculated to be 62.72
$pm/\,^{\circ}\mathrm{C}$ (Fig.~\ref{TOCsil-j3}b).
From this measurement and the fill factors in different regions of waveguide (from Tab.~\ref{tab-fillfactors}) we calculate the TOC of silicon using the following expression:
\begin{equation}\label{eq-tocsil}
\frac{\partial n_{si}}{\partial T} = \frac{\left(\frac{n_g}{\lambda_{res}} \frac{\Delta \lambda}{\Delta T} - \frac{n_g}{n_{ox}} \gamma_{ox}
	\frac{\partial n_{ox}}{\partial T}\right)} {\frac{n_g}{n_{si}}\gamma_{core}}
\end{equation}
This expression is obtained by substituting Eq.~\eqref{eq-dneffdT} in Eq.~\eqref{eq-nefftoc}. In this case, the thermo-optic
coefficient of upper and side cladding (air) is considered to be zero. 
Upon substitution of relevant parameters, we calculated the TOC of silicon as $1.92\times10^{-4}/\,^{\circ}\mathrm{C}$.
While this value appears to be slightly higher than the commonly reported TOC of $1.86\times10^{-4}/\,^{\circ}\mathrm{C}$, we
 note that the TOC itself is a function of temperature. Specifically, Frey. \textit{et. al.} \cite{Frey2006}
have tabulated TOC values of silicon for different temperatures and a range of wavelengths. Based on this reference, we
computed the value of silicon TOC at a mean temperature of 35$\,^{\circ}\mathrm{C}$  (or 308 K)to be $1.93\times
10^{-4}/\,^{\circ}\mathrm{C}$ which agrees well with our measurements. 

\begin{figure}[!htb]
	\begin{center}
		\resizebox{110mm}{!}{\includegraphics *{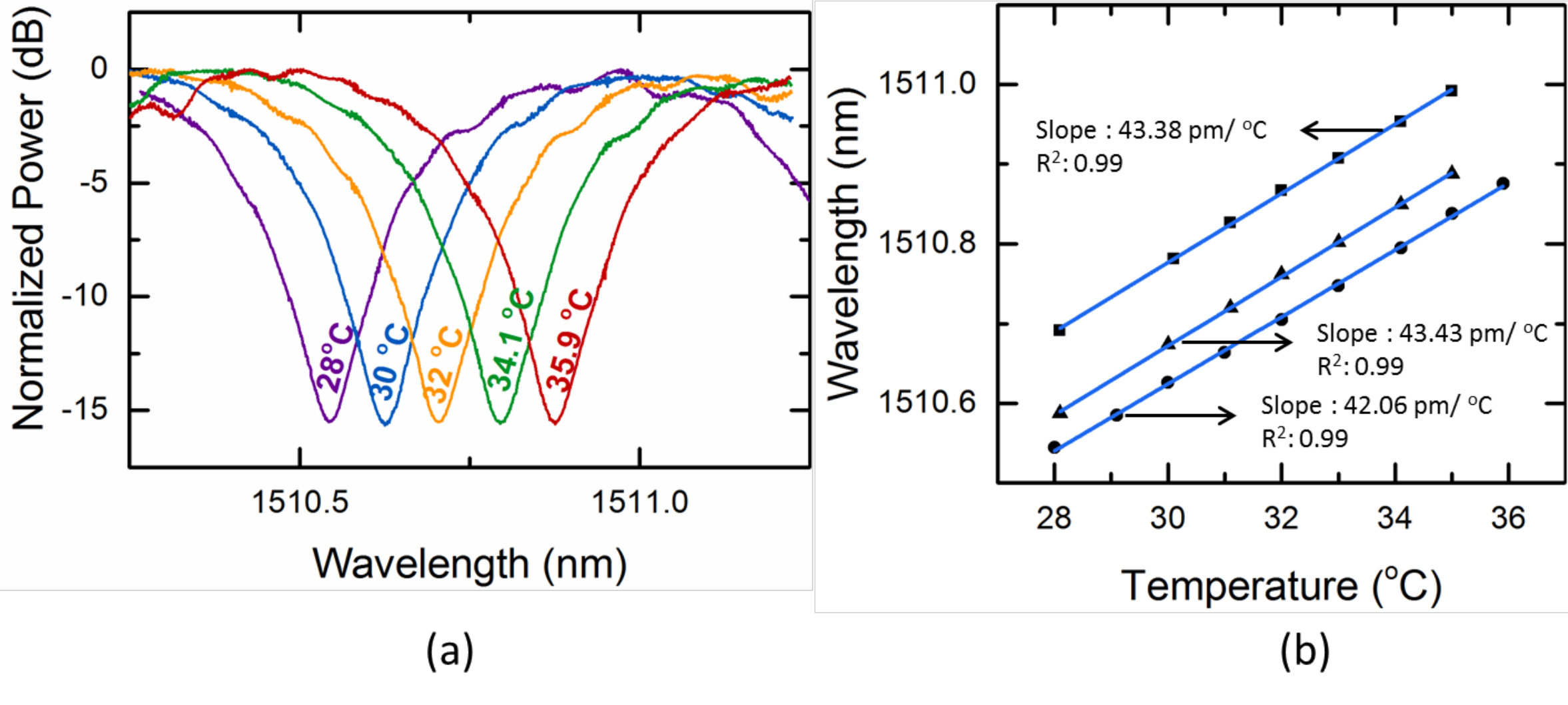}}
		\caption{\small{Measurements for determination of TOC: De-ionized water (a) Positions of a transmission resonance  for different
				 temperatures. Losses caused due to absorption by water (at 1550 nm) result in broader resonance ($\approx 300 pm$)
				 relative to that of air cladding (b) Shift in resonance position as a function of temperature.
			  $R^2$ value for all curve fits exceeded 0.99.}}
		\label{TOCdiw-j3} 
	\end{center}
\end{figure}

To determine the TOCs of liquids, the PDMS reservoir was filled with a small quantity of the analyte and wavelength shift
measurements were performed as described before. TOCs were calculated by substituting wavelength shift slopes $\left(\frac{\Delta
\lambda}{\Delta T}\right)$ and the simulated parameters of Tab.~\ref{tab-fillfactors} in Eq.~\eqref{eq-tocfluid}. 
 The FWHM of resonances (Fig.~\ref{TOCdiw-j3}a)
is noticeably larger than that for air cladding due to absorption of light by water.  
The mean slope (3 trials) of resonance wavelength shift is 42.95  $pm/\,^{\circ}\mathrm{C}$ (Fig.~\ref{TOCdiw-j3}). 
The decreased slope of wavelength shift relative to that of air clad is a result of reduction in the rate of change of
effective index
shift $\left(\frac{\partial n_{eff}}{\partial T}\right)$ owing to the negative thermo-optic coefficient of water.
Using Eq.~\eqref{eq-tocfluid} and simulated parameters of Tab.~\ref{tab-fillfactors} we calculated the TOC of DI-Water to be 
$-1.12\times10^{-4}/\,^{\circ}\mathrm{C}$. This value compares well with those reported previously in literature, as shown in
Tab.~\ref{tab-comparison-toc}.

\begin{figure}[!htb]
	\begin{center}
		\resizebox{110mm}{!}{\includegraphics *{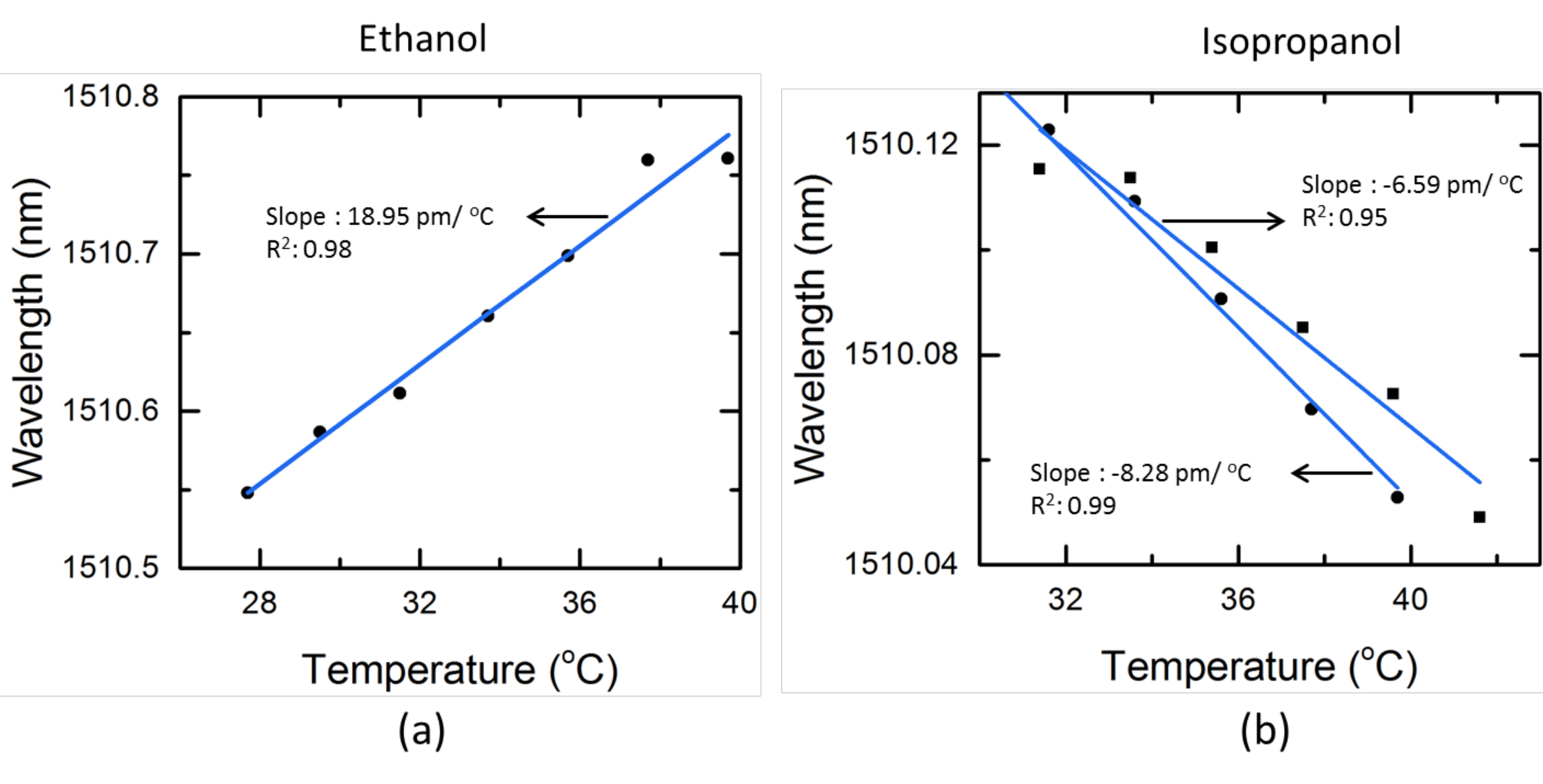}}
		\caption{\small{Measurements for determination of TOC: Organic liquids (a) Resonant wavelength shifts for ethanol cladding. (b) 
		Shifts in resonance wavelength for IPA cladding. In this case, the slope is negative owing to very strong (negative) TOC of
		the analyte.}}
		\label{TOCorg-j3}
	\end{center}
\end{figure}

Next, we measured the TOCs of organic solvents Ethanol and Isopropanol. 
Optical constants of these liquids have been well documented in literature. We limited the experiments with solvents to
a maximum of two trials to avoid disintegration of PDMS reservoir due to swelling \cite{Rumens2015}. 
Between measurements involving different liquids, the chip was cleaned using DI water, dried with nitrogen flow and heated to
remove trace moisture.   
For Ethanol (Fig.~\ref{TOCorg-j3}a), we see that the wavelength shift slope is reduced further to 18.95 $pm/\,^{\circ}\mathrm{C}$
(one trial), owing to stronger negative TOC which is calculated to be $-2.59\times10^{-4}/\,^{\circ}\mathrm{C}$. 
In case of Isopropanol, the mean 
wavelength shift slope (of two trials) is negative at -7.44 $pm/\,^{\circ}\mathrm{C}$. That is, the effective index
shift rate $\left(\frac{\partial n_{eff}}{\partial T}\right)$ of Eq.~\eqref{eq-dneffdT} is negative indicating that
the effect of TOC of cladding liquid outweighs that of core silicon, as seen in Fig.~\ref{TOCorg-j3}b.
Upon substitution of relevant parameters in Eq.~\eqref{eq-tocfluid}, we obtain the TOC of Isopropanol to be 
$-4.21\times10^{-4}/\,^{\circ}\mathrm{C}$.

Thermo-optic coefficients measured using the described method are largely in good agreement with literature reports, as seen
in Tab.~\ref{tab-comparison-toc}, although a slight deviation is observed in case of Ethanol. As mentioned before, we could 
not perform multiple measurement trials with organic solvents owing to the possibility of damage to the PDMS structure. 
Better convergence with literature reports is expected with more measurement iterations using a compatible reservoir
material. 
\begin{table}[!h]
	\begin{center}
		\begin{tabularx}{\textwidth}{|c|l|X|}
			\hline
			Fluid&Our study&Literature Reports with references\\
			\hline\hline
			DI Water&$-1.12\times10^{-4}/\,^{\circ}\mathrm{C}$&$-7.65\times10^{-5}/\,^{\circ}\mathrm{C}$ \cite{Lee2015}, $-1.241\times10^{-4}/\,^{\circ}\mathrm{C}$ \cite{Kim2012}
			\\ \hline
			Ethanol&$-2.59\times10^{-4}/\,^{\circ}\mathrm{C}$&$-3.69\times10^{-4}/\,^{\circ}\mathrm{C}$\cite{Lee2015}, 
			$-3.38\times10^{-4}/\,^{\circ}\mathrm{C}$ \cite{Kim2012}\\ \hline
			Isopropanol&$-4.21\times10^{-4}/\,^{\circ}\mathrm{C}$ &$-4.5\times10^{-4}/\,^{\circ}\mathrm{C}$\cite{Kim2004}, 
			$-4\times10^{-4}/\,^{\circ}\mathrm{C}$ \cite{Qiu2012a}\\ 
			\hline		
			\end{tabularx}
			\caption{\small{Comparison of thermo-optic coefficients determined by our study with values
				reported in literature near 1550 nm wavelength band.}}
		\label{tab-comparison-toc}
	\end{center}
	
\end{table}

It is important to analyze the influence of uncertainties in various simulated and measured parameters on the calculations of
thermo-optic coefficient. Towards this end, 
we analyzed the effect of variation of several critical parameters in Eq.~\eqref{eq-tocfluid} and their effect on the TOCs
of fluids. Major contributions towards uncertainties in TOC calculations arise from discrepancies in the wavelength shift slope
measurements and variations in the width of waveguides used for fill factor simulations. Other sources of deviations in TOC
calculations include uncertainty in the absolute refractive index of liquids used in simulations, and variations in field confinement
factors over the measurement temperature range. However, our analysis showed that the latter two sources are at least
an order of magnitude lower in terms of contribution towards variations in measurements, and were therefore not analyzed further.

\begin{figure}[!htb]
	\begin{center}
		\resizebox{120mm}{!}{\includegraphics *{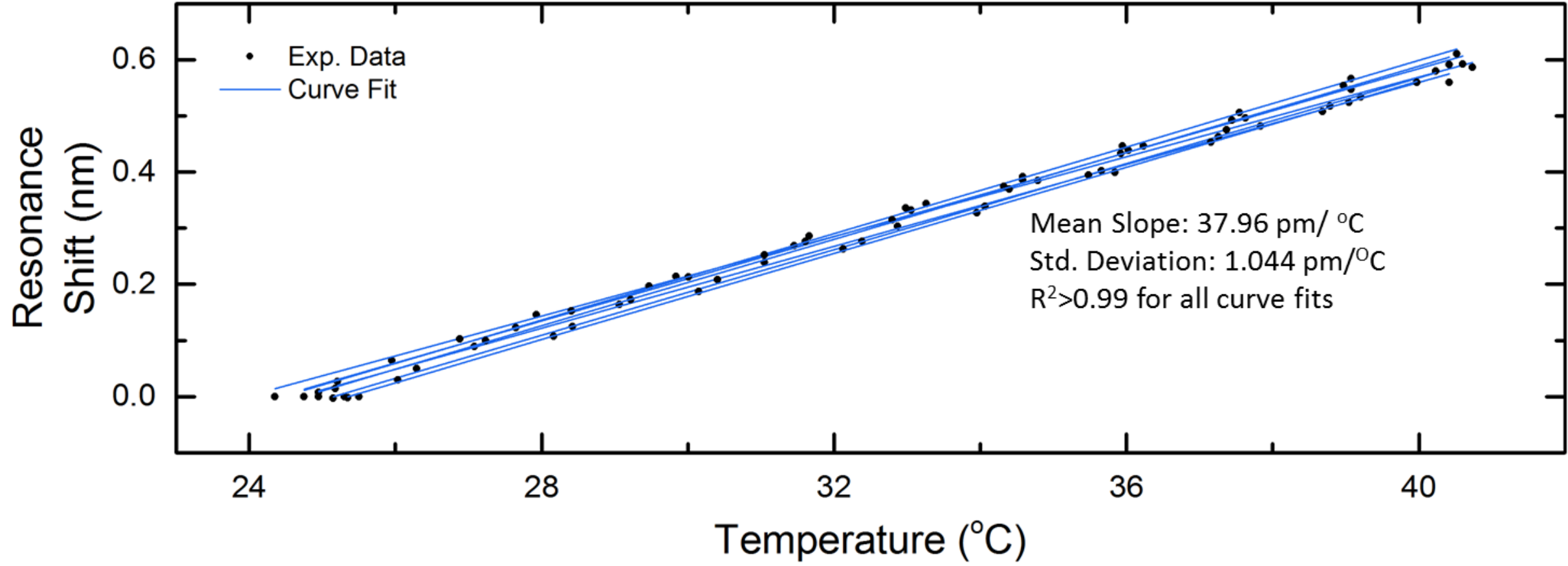}}
		\caption{\small{Determination of error limits of TOC of water. Seven iterations of wavelength shift slope measurements were
				performed.}}
		\label{DIW-LOD-j3}
	\end{center}
\end{figure}

Random variations in slopes of resonance wavelength shifts are a result of various noise affecting the measurement system.  
To obtain an estimate of this variation, we performed seven iterations of slope measurements for the DI-water analyte using the
procedure described earlier. Results of these measurements are plotted in Fig.~\ref{DIW-LOD-j3}.
 The coefficient of determination ($R^2$ value) for curve fits was greater than 0.99 in all trials 
indicating good linearity of wavelength shifts and also the quality of estimated slopes. 
Standard deviation ($\sigma_{\Delta\lambda/\Delta T}$) of seven wavelength shift slopes values was calculated to be 1.044
$pm/\,^{\circ}\mathrm{C}$ .
Upon substitution of Eq.~\eqref{eq-dneffdT} in Eq.~\eqref{eq-tocfluid} and subsequent differentiation of the equation with respect to the slope $\frac{\Delta\lambda}{\Delta T}$ we obtain the following expression. 
\begin{equation}\label{eq-tocerrdlamdT-ch5}
e_{\Delta\lambda/\Delta T}  = \frac{n_{fl}}{\gamma_{fl}} \frac{\sigma_{\Delta\lambda/\Delta T}}{\lambda_{res}}
\end{equation}
This equation translates variations in slope measurements ($\sigma_{\Delta\lambda/\Delta T}$) into equivalent uncertainty in TOC 
estimations which is calculated to be  $e_{\Delta\lambda/\Delta T}=6.42\times10^{-6}/\,^{\circ}\mathrm{C}$.  

Next, we determined the effect of variations in width of waveguide cores on the estimates of TOC. Through measurements on SEM images
taken at different positions on the microring, we computed the mean and standard deviation ($\sigma_w$) of widths to be respectively
387 nm and 4.5 nm. Upon differentiation of Eq.~\eqref{eq-tocfluid} with respect to width of waveguide core, we obtain the following
expression:
\begin{equation}\label{eq-tocerrwdt}
e_{w}  = \Bigg|\frac{n_{fl}}{\gamma_{fl}}\left(\frac{1}{\lambda_{res}} 
\frac{\Delta\lambda}{\Delta T}-\frac{1}{n_{co}}\left(\frac{\partial n_{si}}{\partial T}\right) \frac{\partial \gamma_{co}}{\partial w}\right)
 -\bigg(\frac{n_{fl}}{\gamma_{fl}^2}\frac{\partial\gamma_{fl}}{\partial w}\bigg) \left( \frac{1}{\lambda_{res}}\frac{\Delta\lambda}{\Delta T} - \frac{\gamma_{co}}{n_{co}}\frac{\partial n_{si}}{\partial T}\right)\Bigg| \sigma_{w}
\end{equation}
Here,  $\frac{\partial\gamma_{co}}{\partial w}$ and  $\frac{\partial\gamma_{fl}}{\partial w}$ are the rates of change
of field fill factors in the core  and analyte fluid clads with respect to change in width ($w$) of the core. In deriving
Eq.~\eqref{eq-tocerrwdt}, we have neglected the effect of variation in the buried-oxide fill factor since the contribution
of the term is more than an order of magnitude lower than that due to variations in $\gamma_{fl}$ and $\gamma_{co}$.
Using modal simulations with waveguide core as the variable we found the derivatives of fill factors as
$\frac{\partial\gamma_{co}}{\partial w} = 1.0901/\mu m $ and $\frac{\partial\gamma_{fl}}{\partial w} = -0.625/\mu m$. Substituting
these results in Eq.~\eqref{eq-tocerrwdt} we obtain the contribution towards uncertainty in TOC to
be $e_{w}=5.19\times10^{-6}/\,^{\circ}\mathrm{C}$. The combined effect deviations introduced by slope shift uncertainty 
$e_{\Delta\lambda/\Delta T}$ and core width variations  $e_{w}$ is obtained by computing the 
square root of squares of individual contributions. This value is computed to be
$e_{TOC-DIW} = 8.26\times10^{-6}/\,^{\circ}\mathrm{C}$ which is about 7 \% of the calculated TOC of DI water. If the sensor
for measurement of variations in TOC due to changes in physical/chemical properties of liquids, the detection limit of TOC is desired.
This is taken as three times the standard uncertainty, which for our system computes to be $2.47\times10^{-5}/\,^{\circ}\mathrm{C}$.

\section{Conclusion}
We have reported a method to determine the thermo-optic coefficient of a liquid analyte using silicon photonic microring
resonators. Using the field confinement theory in high contrast waveguides, we have formulated an expression for  describing
the thermo-optic coefficient of a liquid in terms of field confinement factors. Variations in temperature near the
microring is tracked by monitoring electrical resistance of a metal ring structure, enabling real time measurements.
  Experiments for measurement of thermo-optic coefficients of silicon, DI water and organic solvents
Isopropanol and ethanol were described with supporting results. 
We also discussed possible errors in TOC calculations resulting from uncertainties of various parameters. The TOC of water was
determined to be $-1.12\times10^{-4}/\,^{\circ}\mathrm{C}$, with an
uncertainty of $8.26\times10^{-6}/\,^{\circ}\mathrm{C}$. In our experiments, we have used a peltier module for heating the sensor
system. 
However, it is notable that the metal ring used for temperature monitoring can itself be used to heat the analyte for TOC
measurements. We were not successful in achieving this due to electrolysis of polar liquids, once the 
applied voltage to the metal ring exceeded the electrolytic potential. To circumvent this problem, a thin layer of insulating di-electric
can be coated to electrically isolate the metal ring from the analyte liquid. 

\section*{Acknowledgments}
Authors thank the staff of National Nano-Fabrication Centre (NNFC), and Micro and Nano Characterization Facility (MNCF) at the Indian Institute of Science-Bangalore, for their assistance. PRP is grateful to the Ministry of Human Resource Development, Government of India for funding his scholarship. SKS thanks the Ministry of Electronics and Information Technology, Government of India for support.

\end{document}